\documentclass[aps,prb,twocolumn,superscriptaddress]{revtex4-1}
\usepackage{graphicx}
\usepackage{color}
\usepackage{float}
\begin{document}

\title{Probing plasmon-NV$^0$ coupling at the nanometer scale with photons and fast electrons}

\author{Hugo Louren\c{c}o-Martins}
\affiliation{Laboratoire de Physique des Solides, Universit\'e Paris-Saclay, CNRS UMR 8502, F-91405, Orsay, France}

\author{Mathieu Kociak}
\affiliation{Laboratoire de Physique des Solides, Universit\'e Paris-Saclay, CNRS UMR 8502, F-91405, Orsay, France}

\author{Sophie Meuret}
\affiliation{AMOLF, 1098 XG Amsterdam, Netherlands}

\author{Fran\c cois Treussart}
\affiliation{Laboratoire Aim\'e Cotton, UMR 9188 CNRS, Universit\'e Paris-Saclay and ENS Paris-Saclay, Orsay 91405, France}

\author{Yih Hong Lee}
\affiliation{Division of Chemistry and Biological Chemistry, School of Physical and Mathematical Sciences, Nanyang Technological University, Singapore 637371, Singapore}

\author{Xing Yi Ling}
\affiliation{Division of Chemistry and Biological Chemistry, School of Physical and Mathematical Sciences, Nanyang Technological University, Singapore 637371, Singapore}

\author{Huan-Cheng Chang}
\affiliation{Institute of Atomic and Molecular Sciences, Academia Sinica, Taipei 106, Taiwan}

\author{Luiz Henrique Galv\~ao Tizei}
\email{luiz.galvao-tizei@u-psud.fr}
\affiliation{Laboratoire de Physique des Solides, Universit\'e Paris-Saclay, CNRS UMR 8502, F-91405, Orsay, France}

\date{\today}


\begin{abstract}
\textbf{The local density of optical states governs an emitters' lifetime and quantum yield through the Purcell effect. It can be modified by a surface plasmon electromagnetic field, but such a field has a spatial extension limited to a few hundreds of nanometers, complicating the use of optical methods to spatially probe emitter-plasmon coupling. Here we show that a combination of electron-based imaging, spectroscopies and photon-based correlation spectroscopy enables measurement of the Purcell effect with nanometer and nanosecond spatio-temporal resolutions. Due to the large variability of radiative lifetimes of emitters in nanoparticles we relied on a statistical approach to probe the coupling between nitrogen-vacancy centers in nanodiamonds and surface plasmons in silver nanocubes. We quantified the Purcell effect by measuring the nitrogen-vacancy excited state lifetimes in a large number of either isolated nanodiamonds or nanodiamond-nanocube dimers and demonstrated a significant lifetime reduction for dimers.}
\end{abstract}

\maketitle

\textbf{Note:} this manuscript has been published in ACS Photonics: H. Louren\c{c}o-Martins, \textit{et al},\textit{ACS Photonics}, 5, 324 (2018). This is the last version submitted before proof editions.

The radiative lifetime of a light emitter is intrinsically linked to the local density of optical states (LDOS), $\rho(\vec{r},\omega)$ ($\vec{r}$ represents position in space and $\omega$ frequency). The presence of an optical cavity or a plasmonic structure leads to the increase of the LDOS which can be quantified by a transition rate enhancement factor, $\gamma = \rho(\vec{r},\omega)/\rho_{0}(\vec{r},\omega)$\cite{Lukas} ($\rho$ and $\rho_0$ being the LDOS with and without the a cavity or plasmonic structure), what is known as Purcell effect \cite{Purcell1946}. Owing to the Fermi golden rule, the radiative lifetime of an emitter interacting with these photonic structures is modified by $\gamma$. Due to the short lifetime of surface plasmons (SP) \cite{Bosman2013} (typically few femtoseconds), their coupling with quantum emitters (QE) is usually weak. In this regime, the Purcell effect is predominant.  

To date, only two strategies have been considered to measure the Purcell effect in QE-SP dimers. From a microscopic perspective, experiments were carried out on a single dimer \cite{Farahani2005, Kuhn2006, Schietinger2009, Beams2013} with a precise positioning of the QE. Alternatively, macroscopic measurements were performed simultaneously on a large ensemble of dimers \cite{Aberra2012, Fedutik2007}. Both approaches present limitations due to the large variability of the isolated QEs lifetimes \cite{Fisher2004, Storteboom, Lim2009,Andersen2016}.

Here, to overcome this intrinsic variability we adopt a statistical method applied to individual nano-objects, where the lifetime of large sets of isolated QEs and dimers are measured. In our experiments,  neutral nitrogen-vacancy (NV$^0$) centers in nanodiamonds dipoles were coupled to surface plasmons (SP) on Ag nanocubes. To quantify the Purcell effect at the nanoscale we applied a combination of electron-based imaging, spectroscopies and photon-based correlation spectroscopy that allowed us to achieve the required spatial and temporal resolutions with large throughput. 

\begin{figure},
    \includegraphics[width=\columnwidth]{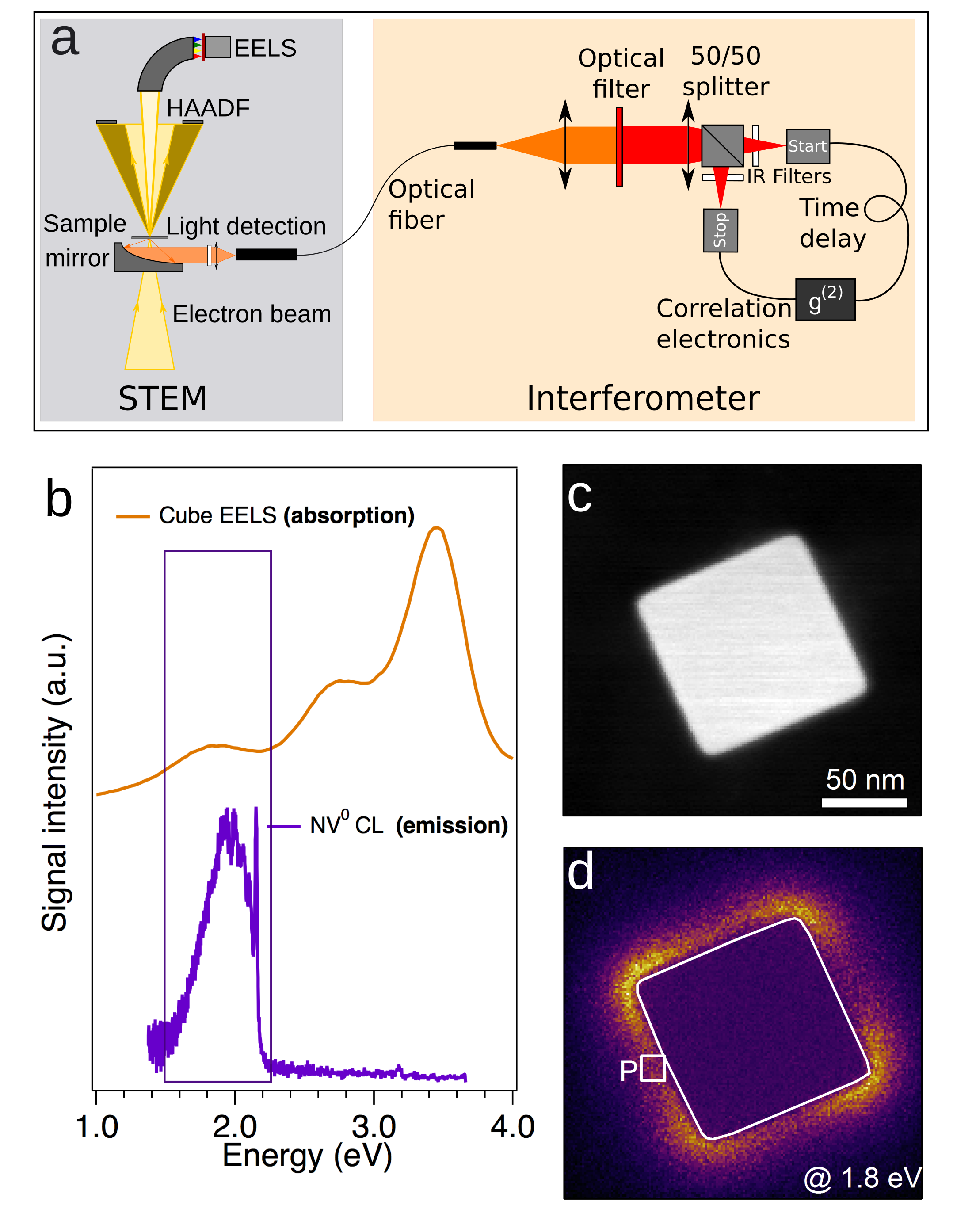}
    \caption{(a) Schematic of the STEM and HBT interferometer setup. (b) EELS spectrum (absorption) of a Ag nanocube measured in region $P$ indicated in (d). Each peak corresponds to a plasmon resonance. The purple rectangle indicates the energy window where the optical filter is active. (c) ADF image of a Ag nanocube. (d) Energy filtered (at 1.8 eV) EELS map of the silver nanocube showed in (c). Scale bar: 50 nm.}
    \label{FIG1}
\end{figure}

Modern scanning transmission electron microscopes (STEM) are versatile, allowing the use of nanometer-wide electron probes to obtain complementary information from spectroscopic, diffraction and imaging techniques. Specifically electron energy loss spectroscopy (EELS) and cathodoluminescence (CL) have shown their remarkable capability to measure absorption and emission spectra with nanometer spatial resolution \cite{Nelayah2007, Zagonel2011, Losquin2015, Kociak2017}. Figure 1 shows a typical EEL spectrum for an isolated Ag nanocube and a CL spectrum for an isolated nanodiamond containing NV$^0$ centers. The EEL spectrum presents three peaks corresponding to different plasmon modes. The peak at ~1.8 eV with a ~500 meV width matches the energy range of the NV$^0$ emission, indicating the possibility of coupling between SP and the NV$^0$ centers. We note that SP spatial intensity distribution is highly anisotropic, as seen in Fig. \ref{FIG1}(c-d) for the mode at 1.8 eV. In principle, this could play a role into where the nanodiamond should be positioned. As discussed later, this does not play a major role due to the large spatial extent of SP modes.

Lifetimes of isolated nanodiamonds and nanodiamond/nanocubes dimers have been inferred from the second order correlation function (g$^{(2)}(\tau)$) of light emitted from NV$^0$ centers excited by the fast electrons, as already demonstrated for different systems \cite{Meuret2015, Meuret2016, Meuret2017, tizei2017chapter}. In short, each electron excites more than one electron-hole pair above the bandgap of diamond. These pairs excite within a few picoseconds more than one localized defect. The end result is the emission of more than one photon within the emitter's lifetime, leading to the effective formation of a light pulse and a second order correlation function presenting a bunching behavior. This bunching peak has a time decay constant equal to the emitter's total lifetime, allowing its measurement. Note that lifetimes can also be measured using a pulsed electron source and a cathodoluminescence setup, but with limited spatial resolution (50 nm) up to now \cite{Merano2005}. g$^{(2)}(\tau)$ was measured using a Hanbury Brown and Twiss (HBT) interferometer (see Fig. \ref{FIG1}(a)) coupled to the light collection system \cite{Tizei2013}. Annular dark field (ADF) images allows us to determine the position of the nanodiamond with respect to the nanocube and the nanocube size with a nanometer resolution, giving us all the parameters necessary to estimate expected the enhancement factor.  

\begin{figure}
    \includegraphics[width=\columnwidth]{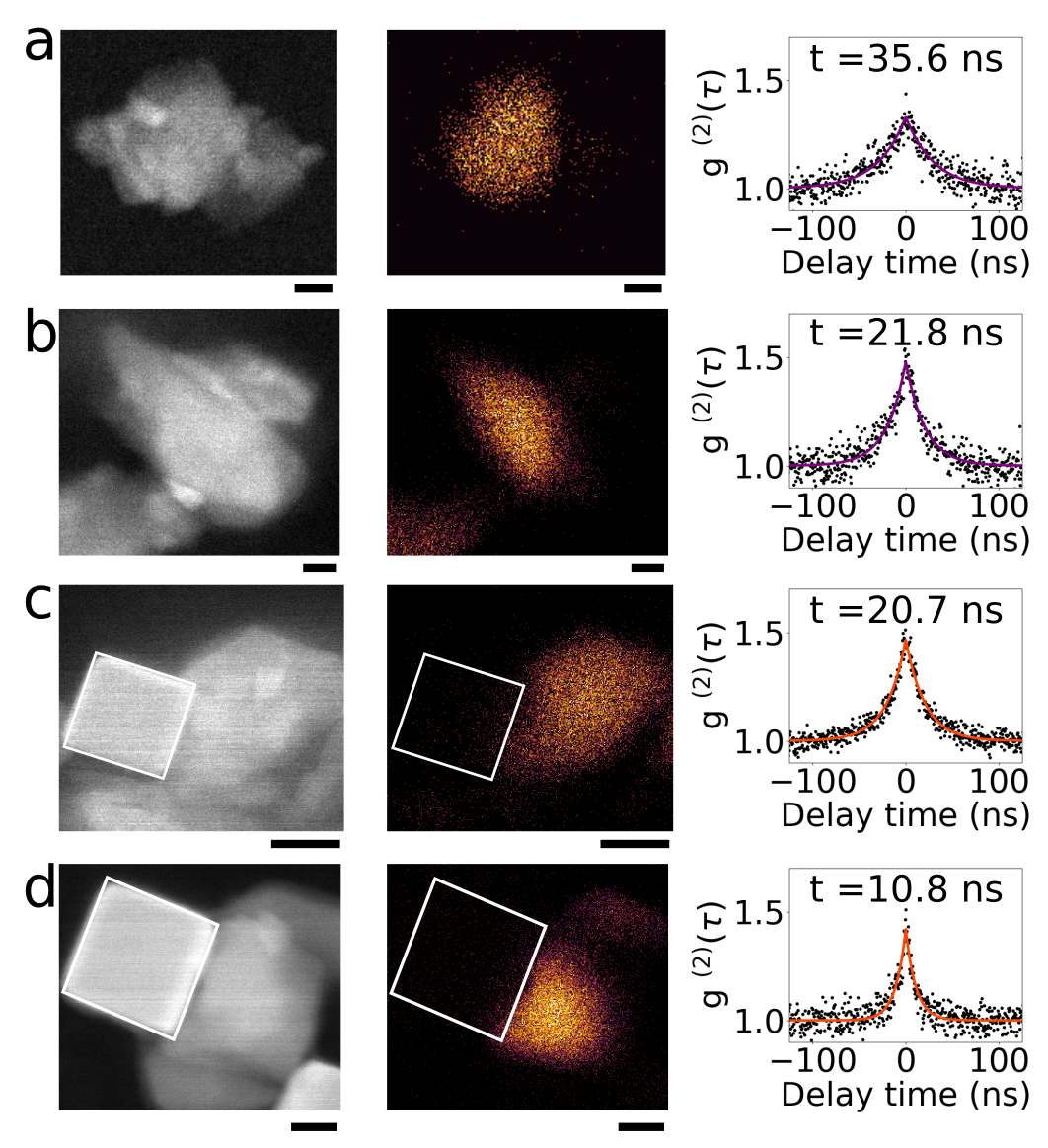}
    \caption{(a-b) From left to right: ADF image, energy-filtered NV$0$ emission intensity image and g$^{(2)}$ correlation function of single nanodiamonds. (c-d) From left to right: ADF image, energy filtered CL maps (see Fig. \ref{FIG1}) and g$^{(2)}$ correlation function of nanodiamonds close to a silver nanocube. Scale bars: 50 nm.}
    \label{FIG2}
\end{figure}

Samples were prepared by sequentially drop casting solutions of Ag nanocubes and nanodiamonds containing multiple NV$^0$ centers onto a 15 nm thick Si$_3$N$_4$ membrane. Nanodiamonds typically appear as aggregates (Fig. \ref{FIG1}a). However, among these aggregates a larger and luminescent nanoparticle is always observed, which is the one considered in each measurement.  Most of the nanodiamonds tend to be adsorbed on the nanocubes faces. Moreover, a distribution of sizes is observed in our nanocubes sample (100 nm average size with some variation). This influences the energy of a specific SP. But as a continuum of SP modes is observed for each given nanocube and the NV$^0$ emission is spectrally wide, coupling is always possible. Isolated nanodiamonds and dimers were identified using ADF images that are acquired simultaneously with wavelength filtered CL maps (first and second columns in Fig. \ref{FIG2}). Quick access to this information allows an effortless identification. After a target isolated nanodiamond or dimer is selected, the g$^{(2)}(\tau)$ function of the emitted photons is measured using the HBT interferometer while the electron beam scans a fixed small area on the nanodiamond. The photon counting  rate and the ADF image can be recorded live, allowing sample drift to be corrected by repositioning the scanning area. In total, the lifetime of 56 isolated nanodiamonds and 62 dimers were measured in the same sample in a single experimental run, ensuring identical experimental conditions.

Examples of measurements in two isolated nanodiamonds and two dimers are shown in Fig. \ref{FIG2}(a-b) and Fig. \ref{FIG2}(c-d), respectively. The isolated nanodiamond in Fig. \ref{FIG2}(a) and the dimer in \ref{FIG2}(d) have lifetimes $36\pm 5$ ns and $\tau=11\pm 1$ ns, respectively, in agreement with an enhancement effect. However, isolated nanodiamonds and dimers (Fig. \ref{FIG2}(b-c)) with similar lifetimes are also present ($\tau=22\pm 2$ ns and $\tau=21\pm 1$ ns in these examples). Such observations occur due to the large dispersion in NV$^0$ lifetimes in nanodiamonds already reported in the literature \cite{Storteboom, Tisler, Beveratos}. The lifetime for NV$^0$ in bulk diamond is 19~ns~\cite{Liaugaudas} and it is distributed between 10 and 40~ns~\cite{Storteboom} in nanodiamond. A similar behavior is known for the NV$^-$  (charged NV) center, for which the lifetime changes from 13~ns in bulk~\cite{Collins1983} to larger values (17~ns and 25~ns) in nanoparticles~\cite{Tisler, Beveratos}, with a broad distribution \cite{Inam:2011fq}. The excited state lifetime increase is the result of the smaller LDOS in nanoparticle than in bulk (i.e. $\gamma<1$, as predicted in the electrostatic regime of particle size smaller than the vacuum emission wavelength of the embedded emitter), while the lifetime dispersion is mainly due to nanoparticle size variability~\cite{Greffet2011}.
 
To overcome this variability, a measurement of the Purcell effect can be performed by sequentially probing the NV$^0$ lifetime of an isolated nanodiamond, followed by coupling the nanodiamond to a plasmonic structure (either by mechanical movement or lithography), as shown by Beams \textit{et al} \cite{Beams2013}. However, this approach would normally involves taking the sample out of vacuum and performing a series of processes which can modify its local environment and, hence, its lifetime. Here we overcome the variability by measuring the excited state lifetime in a large ensembles of either isolated nanodiamonds or dimers.
     
\begin{figure}
    \includegraphics[width=\columnwidth]{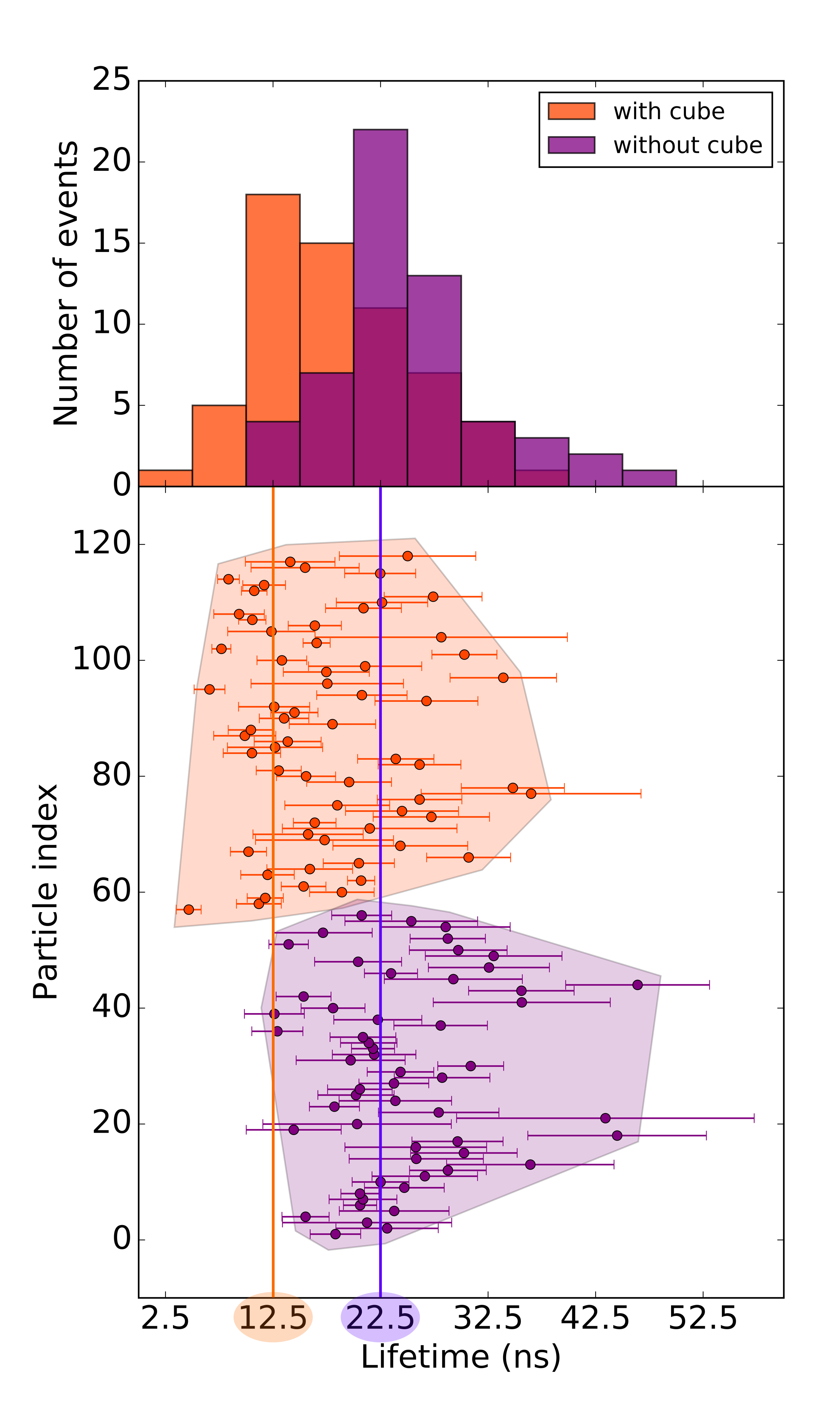}
    \caption{Distribution of nanodiamonds cathodoluminescence lifetime alone (purple) and close to a silver nanocube (orange) measured on a population of 118 diamonds. The two vertical lines indicate the maxima of the nanodiamonds' lifetime distribution alone (22.5$\pm$2.5 ns in purple) and close to a silver nanocube (12.5$\pm$2.5 ns in orange). }
    \label{FIG3}
\end{figure}

The histograms (top) and a scattered plot (bottom) of the lifetime of isolated nanodiamonds (purple) and dimers (orange) are shown in Fig. \ref{FIG3} (see SI for the complete data set). The average lifetimes of isolated nanodiamonds and dimers are $24\pm 5$ ns and $18\pm 4$ ns (the most probable values are $22.5\pm 2.5$ ns and $12.5\pm 2.5$ ns). The two distributions overlap. However, they are significantly distinct, as confirmed by the Wilcoxon-Mann-Whitney statistical u-test ($p=1.91\times 10^{-6}$). Therefore, the 40\% reduction of lifetimes can be unambiguously associated to a Purcell effect with a spontaneous decay rate enhancement factor of 1.4, indicating some coupling, although weak, between NV$^0$ centers in nanodiamonds and SP in Ag nanocubes.

Lifetimes shorter than that in bulk are observed in both histograms, although with a small probability. These occur due to non-radiative decay channels involving e.g. surface defects, which do affect the excited state lifetime. Overall, the coupling leads to a rigid shift of NV$^0$ lifetime histogram to shorter values. These features are only accessible with a statistical approach which emphasizes the strength of our experimental method.

As recently pointed out \cite{Meng2017}, the calculation of the luminescence enhancement of defects in presence of a plasmonic field is an intricate problem. A quantitative simulation of our experiment would require precise knowledge of a large set of parameters: the shape of the nanodiamonds, exact number of defects, their position in the nanoparticle, their respective lifetime and the profile of the plasmonic field within the nanodiamond. Although we do not tackle this problem in the current paper, we have performed numerical calculations to verify that the order of magnitude of the expected effect matches our observations. A key point for these calculations is the presence of numerous NV$^0$s in our nanodiamonds. In principle, the NV$^0$ lifetime may vary with the position of the emitter within a nanoparticle for subwavelength-sized nanodiamonds (electrostatic regime) but experimentlly we have observed that it is constant throughout the nanodiamond, as pointed out by Greffet \textit{et al}\cite{Greffet2011}. No variations within the same nanodiamond was observed, despite the nanometer spatial resolution provided by our experiments. Even if electron-hole diffusion in the nanodiamond could decrease our expected spatial resolution (as it is known to occur \cite{Tizei2012}), variations of NV$^0$s lifetime could be observed if they occurred in scales larger than ~50 nm.

\begin{figure}
    \includegraphics[width=\columnwidth]{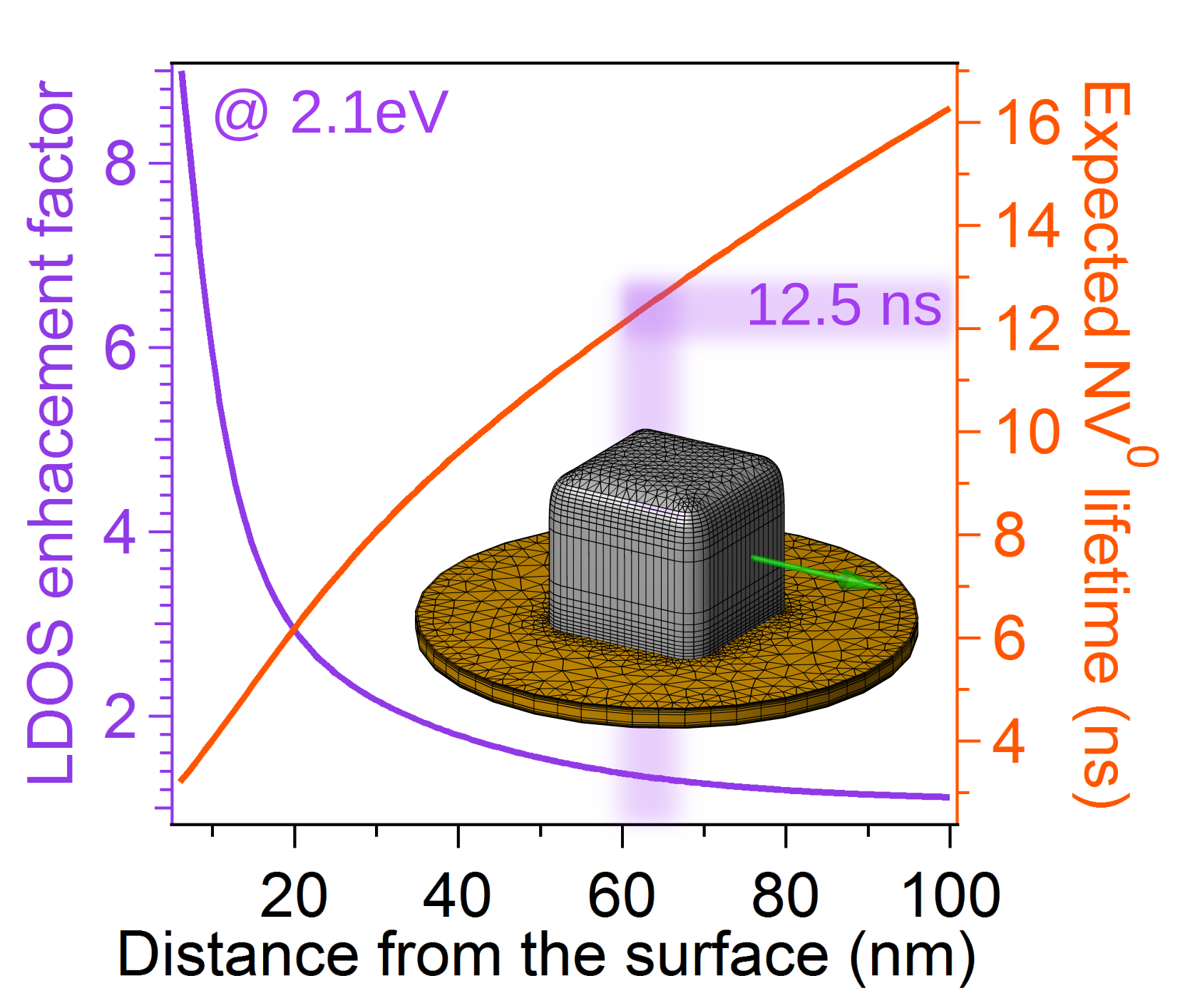}
    \caption{(Purple) LDOS enhancement factor , $\gamma$ due to SP sustained by an Ag cube, at 2.1 eV calculated along the green arrow represented on the inset. (Orange) Corresponding expected NV$^0$ lifetime. When the LDOS enhancement factor is equal to one, the lifetime is assumed to 22.5 ns (as measured in Fig. 3).}
    \label{FIG4}
\end{figure}

The expected radiative rate enhancement factor due to the Purcell effect was calculated by considering an isolated Ag nanocube, taken the Si$_3$N$_4$ substrate into account. Calculations performed with the \texttt{MNPBEM} toolbox \cite{Hohenester}  show a SP mode centered at 2.1 eV, in agreement with the experimental value (Fig. \ref{FIG1}(a)). Considering that most nanodiamonds were observed on a facet of the nanocubes, we calculated the LDOS enhancement factor at 2.1 eV along a line centered on a nanocube face and perpendicular to it, for an emitter at a distances between 10 to 100 nm from the nanocube surface (arrow on Fig. \ref{FIG4}). 

The enhancement factor is plotted on Fig \ref{FIG4} (purple). Taking into account the most probable lifetime for NV$^0$ in isolated nanodiamonds is $22.5$ ns, we plotted in orange the expected NV$^0$ lifetime given the calculated enhancement factor. We see that the most probable lifetime measured in presence of a nanocube ($12.5$ ns) corresponds to a distance of $65$ nm, also associated to the enhancement factor 1.4. This value is in qualitative agreement with the possible distance of a NV$^0$ center to the surface of a 100-200 nm nanodiamond (sizes typically observed in our sample). Although this result relies on the specific positioning of the emitters in the dimer, obtaining a consistent distance value is a strong evidence to support our conclusion.

We have used a combination of fast electron/photon techniques to quantify the Purcell effect resulting from the coupling of dipolar emitters embedded in nanoparticles to plasmonic structures. NV$^0$ in nanodiamond-SP coupling is evidenced by the reduction of the mean excited state lifetime of a distribution of individual isolated nanodiamonds and dimers measurements. This effect could have been masked if we had limited our study to a few objects because of the instrinsic lifetime dispersion. We have shown that combination of fast electron/photon techniques provide the required measurement throughput, spatial and temporal resolutions to disentangle the two effects. The ensemble of techniques described here can be applied seamlessly to any emitter with excited state lifetime in the $0.5$ ns - $50$ ns range which emit light under electron irradiation, covering a wide range of nanoscale systems.
\\
\\

\noindent \textbf{\textit{Acknowledgements}}\\
This work was financially supported by the National Agency for Research under the program of future investment TEMPOS-CHROMATEM with the reference ANR-10-EQPX-50.
\\

\noindent \textbf{\textit{Methods}}\\
Experiments have been performed with a Vacuum Generator HB-501 STEM equipped with a cold field emission electron gun operating at 60 kV, a Gatan 666 EEL spectrometer and the Attolight M\"onch 4107 Cl-system. An homemade HBT interferometer has been implemented on the nano-CL system  (Fig \ref{FIG1}(a)). The stage of the microscope is cooled with liquid nitrogen down to 150 K. EEL spectrum images were deconvolved using a Richardson-Lucy algorithm \cite{Gloter2003}, with a zero-loss peak (ZLP) width reduction from 0.33eV to 0.1eV. Lifetimes were measured as described by Meuret \textit{et al}\cite{Meuret2015, Meuret2016}, allowing measurements in a few tens of seconds.
Nanocubes were synthesized via a polyol reduction route \cite{Tao2006}. 
The g$^{(2)}(\tau)$ measurements were performed with sampling time of 512 ps, using $\tau$-SPADs single photon avalanche photodiodes and a PicoHarp 300 from PicoQuant. The overall system has a response time of 130 ps.

\bibliography{CubeNV0.bib}

\end{document}